\documentclass{ws-ijmpe}
\usepackage[super,compress]{cite}
\begin{document}

\markboth{R.V. Jolos, E.A. Kolganova}{Collective model with isovector pair and alpha-particle type correlations}

\catchline{}{}{}{}{}

\title{Collective model with isovector pair and alpha-particle type correlations}

\author{R. V. Jolos\footnote{Corresponding author}, E. A. Kolganova}
\address{Bogoliubov Laboratory of Nuclear Physics\\
Joint Institute for Nuclear Research, 141980 Dubna, Russia,\\
Dubna State University, 141982 Dubna, Russia\\
*jolos@theor.jinr.ru}
\author{D. A. Sazonov}
\address{Dubna State University, 141982 Dubna, Russia}

\maketitle

\begin{history}
\end{history}

\begin{abstract}

The collective Hamiltonian including isovector pairing and $\alpha$-particle type correlation degrees of freedom is constructed. The Hamiltonian is applied to description of the relative energies of the ground states of even-even nuclei around $^{56}$Ni. A satisfactory description of the experimental data is obtained. A significant improvement of the agreement with the experimental data compared to our previous calculations is explained by inclusion in the Hamiltonian of the dynamical variables describing $\alpha$-particle type correlations.

\end{abstract}


\ccode{PACS numbers: 21.10.-k, 21.10.Dr, 21.60.Ev}


\section{Introduction}

It is well known how important is the role of the pair correlations of nucleons in description
of the structure of atomic nuclei \cite{Bohr1,Belyaev1,Soloviev1,Zelevinsky1}. In the case of nuclei with
$N\approx Z$ in which protons and neutrons fill the same single particle levels it is
necessary to take into account not only pair correlations of nucleons of the same kind
but also neutron-proton pair correlations. It means that at least the isovector pair correlations should be considered.

The isovector pair correlations were first considered by A.Bohr \cite{Bohr2} and continued in Refs.
\cite{Dussel1,Jolos1} In the last years a lot of papers have been dedicated to this problem.
The revival of the interest to this problem is related to the open experimental opportunity
to study structure of the heavier nuclei close to the $N=Z$ line where the isovector pair correlations play
an important role. Another important subject related to the isovector pair correlations is
the masses of nuclei \cite{Neergard,Bently}.

In principle, in nuclei with $N\approx Z$ both isovector $T=1$ and isoscalar $T=0$ pair correlations can take place.
The majority of investigations have been concentrated on the study of the proton-neutron
pair correlations in the framework of the BCS-type microscopic models
\cite{Goswami1,Camiz1,Camiz2,Goswami2,Chen,Goodman1,Goodman2,Goodman3}. It was noted also
that it is important in these type of studies to conserve exactly the particle number and isospin
\cite{Zeng,Engel1,Engel2,Dobes,Satula}. The approach keeping a conservation of isospin and
based on the introduction of the alpha-particle like quartets was developed in
\cite{Sandulescu1,Sandulescu2,Sandulescu3}.

In our previous paper \cite{Jolos2} isovector pair correlations have been considered in the
framework of the collective model formulated in Refs.\cite{Dussel1,Jolos1}. This approach was
applied to consideration of the energies of the ground states of nuclei around $^{56}$Ni.
The results of calculations have shown that especially large deviations from the experimental
data are obtained for the ground states of nuclei with even numbers of $Z$ and $N$, i.e.
for nuclei which can be considered as systems of some numbers of $\alpha$-particle type quartets. This indicates that isovector pairing
forces although contribute but don't exhaust the alpha-particle type correlations in considered nuclei. For this
reason, the aim of the present paper is to construct the collective Hamiltonian which
includes not only the isovector pairing degree of freedom but also those degrees of freedom
which could be responsible for the additional alpha-particle type correlations. This
Hamiltonian is applied below to description of the relative energies of the ground states
of nuclei around $^{56}$Ni. The expressions for the amplitudes of the two-nucleon and
$\alpha$-particle transfer reactions are also presented.

\section{Collective Hamiltonian. Kinetic energy.}

There are two sets of dynamical variables, namely,  corresponding to pair addition and pair
removal modes, needed to describe pair correlations in nuclei \cite{Bohr2}. Each of these two
sets of variables is characterized by the three projections of isospin corresponding to
neutron-neutron, neutron-proton and proton-proton correlated pairs. Thus, there are  six
collective variables in total needed to describe isovector pair correlations. They can be presented by the complex isovector $z_{1\mu}$ ($\mu=0,\pm 1$). It was suggested in \cite{Dussel1} to separate in $z_{1\mu}$
the variables related to isospin invariance and gauge invariance
\begin{eqnarray}
\label{eq1}
z^+_{1\mu}=\Delta\exp(\imath\phi)\left(D^1_{\mu 0}(\psi_1,\psi_2,\psi_3)\cos\theta \phantom{\frac{1}{\sqrt{2}}}\right.\nonumber\\
\left.+\frac{1}{\sqrt{2}}(D^1_{\mu 1}(\psi_1,\psi_2,\psi_3)+D^1_{\mu -1}(\psi_1,\psi_2,\psi_3))\sin\theta\right).
\end{eqnarray}
Here $D^1_{\mu 1}(\psi_k,\psi_2,\psi_3)$ is the Wigner function and $\psi_1,\psi_2,\psi_3$
are Euler angles characterizing orientation in isospace. The angle $\phi$ is conjugate to the
operator of the number of the nucleon pairs ${\hat N}=-\imath\partial/\partial\phi$
added or removed from the basic nucleus. The variable $\Delta$ characterizes
a strength of the isovector pair correlations and $\theta$ describes isospin structure
of the pair correlations in the intrinsic frame.

In addition to $z_{1\mu}$ we introduce a dynamical variable $\alpha$ describing the
$\alpha$-particle type correlations
\begin{eqnarray}
\label{eq2}
\alpha=\exp({2\imath\phi})a.
\end{eqnarray}
This variable depends on the same gauge angle $\phi$, however, with factor 2 in the exponent
since addition or removal of the $\alpha$-particle type quartet changes the number
of nucleon pairs in the nucleus by two. The variable $a$ characterizes a strength of the $\alpha$-particle
type correlations.
This dynamical variable has no isospin indices
since it corresponds to excitation of the isoscalar mode.

Our next task is to construct  a collective Hamiltonian depending on
$z_{1\mu}$ and $\alpha$. The greatest challenge is a construction of the kinetic
energy operator. We begin by writing a classical expression for the kinetic energy
\begin{eqnarray}
\label{eq3}
T=\frac{1}{2}B_{\Delta}\sum_{\mu}\dot{z}^*_{1\mu}\dot{z}_{1\mu}+\frac{1}{2}B_{\alpha}\dot{\alpha}^*\dot{\alpha}
\end{eqnarray}
It is assumed in (\ref{eq3}) that inertia coefficients do not depend on dynamical variables. This
Hamiltonian can be quantized using Pauli prescription \cite{Pauli}. The procedure is analogous to
that used in \cite{Greiner} and for the isovector pairing mode is presented in Ref.\cite{Dussel1}.  We extended it by including the $\alpha$-particle correlation mode which is very
important for the correct description of the energies as it will be shown below. The details
of derivation are given in the Appendix.

The expression for the kinetic energy term of the collective Hamiltonian containing both the isovector pairing
and isoscalar $\alpha$-particle type correlation degrees of freedom is given by expression (\ref{eqa22}).

The collective
potential for isovector pair correlations depends on two invariants: $\Delta^2$ and $\Delta^4\cos^22\theta$.
In \cite{Jolos4} collective potential for isovector pair correlations was derived using the boson representation technique.
It was shown there that the minimum of the potential is located at $\theta$=0. We assume below that the  amplitude of fluctuations of $\theta$
around $\theta$=0 is small for the lowest states with given $N$ and $T$. As a consequence we put $\theta$=0 in the expression
for the kinetic energy term in (\ref{eqa22}) if it does not create singularities. In the case of the kinetic energy of $\theta$ vibrations
and the moment of inertia of the isospin rotations around $z$-axes we keep the lowest order terms in $\theta$.
 As a result we obtain for ${\hat T}$:
\begin{eqnarray}
\label{eq4}
{\hat T}=\frac{1}{2}\frac{{\hat T}_x^2+{\hat T}_y^2}{B_{\Delta}\Delta^2}+\frac{1}{2}\frac{{\hat N}^2}{B_{\Delta}\Delta^2+4B_{\alpha}a^2}\nonumber\\
-\frac{1}{2B_{\Delta}\Delta^4\sqrt{B_{\Delta}\Delta^2+4B_{\alpha}a^2}}\frac{\partial}{\partial\Delta}\Delta^4\sqrt{B_{\Delta}\Delta^2+4B_{\alpha}a^2}\frac{\partial}{\partial\Delta}\nonumber\\
-\frac{1}{2B_{\Delta}\Delta^2}\frac{1}{\theta}\frac{\partial}{\partial\theta}\theta\frac{\partial}{\partial\theta}+\frac{{\hat
T}_z^2}{B_{\Delta}\Delta^2\theta^2}\nonumber\\
-\frac{1}{2B_{\alpha}\sqrt{B_{\Delta}\Delta^2+4B_{\alpha}a^2}}\frac{\partial}{\partial a}\sqrt{B_{\Delta}\Delta^2+4B_{\alpha}a^2}\frac{\partial}{\partial a}
\end{eqnarray}

As the next step we decouple approximately $\theta$ and $\Delta$ modes replacing $\Delta$ in the kinetic energy term for $\theta$-vibrations
by the average value of $\Delta^2$, namely, by $\langle \Delta^2\rangle$. This is done in analogy with the consideration
of the critical point symmetries in nuclear structure physics related to the collective
quadrupole excitations \cite{Iachello1,Iachello2,Iachello3}. For the lowest states with given $N$ and $T$ we
can put $T_z$=0. As the result, the $\theta$-dependent part of the collective Hamiltonian, being completely separated from the other terms,
will give the same contribution into the energies of the lowest states  for each $N$ and $T$.
Since we will compare with the experimental data only the relative energies of nuclei counted from the energy
of the ground state of the basic nucleus, we exclude $\theta$-mode from consideration below.

\section{Collective Hamiltonian. Potential energy}

In our previous paper \cite{Jolos2} we based on the phenomenological approach to determination of the collective potential and considered various variants of a $\Delta$-dependence of the collective potential.
It was found that Davidson's potential containing one dimensional and one nondimensional parameters is the best suited
to describe the energies of the  collective pair excitations of nuclei in the $^{56}$Ni region. By changing the value
 of the nondimensional parameter we can describe all intermediate cases between vibrational and rotational
 limits of pair correlations. We have seen in the previous section that
 inclusion into the Hamiltonian of the collective variable corresponding to the $\alpha$-particle mode leads to a
 significant complication of the terms in ${\hat T}$ corresponding to kinetic energies of $\Delta$- and $a$- modes. There appears the expression $\sqrt{B_{\Delta}\Delta^2+4B_{\alpha}a^2}$ which changes the coefficients
 at the first derivatives over $\Delta$ and $a$. However, it is well known that by some transformation of the wave function it is possible to change a coefficient at the first derivative or even to eliminate this term in the Hamiltonian.
  It leads to appearance of additional terms
 in the expression for the potential energy. But in our case, when the potential energy is treated phenomenologically,
 we can ignore this effect and  exclude from consideration the corresponding factors in the expression for the
 kinetic energy.
 As the result, the Hamiltonian takes the following form where we put $T(T+1)$ instead of $({\hat T}_x^2 + {\hat T}_y^2)$
 since for the lowest states for each set of $N$ and $T$, $T_z$=0:
 \begin{eqnarray}
\label{eq5}
{\hat H}=\frac{T(T+1)}{2B_{\Delta}\Delta^2}+\frac{N^2}{2B_{\Delta}\Delta^2+4B_{\alpha}a^2}
-\frac{1}{2B_{\Delta}\Delta^5}\frac{\partial}{\partial\Delta}\Delta^5\frac{\partial}{\partial\Delta}\nonumber\\
-\frac{1}{2B_{\alpha}}\frac{\partial^2}{\partial a^2}+\frac{1}{2}C_{\Delta}\left(\frac{\Delta^4_0}{\Delta^2}+\Delta^2\right)
+V_{\alpha}(a)
\end{eqnarray}
In (\ref{eq5}) $V_{\alpha}$ is the potential energy of the $\alpha$-mode.

To determine dependence of the potential energy on $a$ let us consider the relative energies of nuclei around $^{56}$Ni
which differ by some number of $\alpha$-particles, i.e. $^{48}$Cr, $^{52}$Fe, $^{60}$Zn, and $^{62}$Ge. In \cite{Bohr2,Bes1}
there was suggested a procedure to extract from the experimental data the quantities which can be directly compared with the results
of calculations based on the collective Hamiltonian. It is necessary to do since the ground state energies of different nuclei
are compared. This procedure has been used in Ref\cite{Jolos3}.
According to this procedure it is necessary to subtract from the experimental binding energies of nuclei
under consideration those contributions  that are generated by the sources other than the isovector monopole pairing and
$\alpha$-particle type correlations. For nuclei around the basic nucleus with $A=A_0$ and $Z=Z_0$, which in our case is
$^{56}$Ni, the following quantity is defined.
\begin{eqnarray}
\label{eq6}
E(A,Z)=-\left(B_{exp}(A,Z)-B_{LD}(A,Z)\right)+\left(B_{exp}(A_0,Z_0)-B_{LD}(A_0,Z_0)\right),
\end{eqnarray}
where $B_{exp}(A,Z)$ is the experimental binding energy of the nucleus with mass number $A$ and charge $Z$.
The quantity $B_{LD}(A,Z)$ is defined by the liquid drop mass formula without the symmetry energy and
the pairing energy terms:
\begin{eqnarray}
\label{eq7}
B_{LD}(A,Z)=a_vA-a_sA^{2/3}-a_c\frac{Z(Z+1)}{A^{1/3}},
\end{eqnarray}
where $a_v$=15.75 MeV, $a_s$=17.8 MeV, and $a_c$=0.711 MeV.

Fig. 1 shows the spectrum of the ground state energies of nuclei that differ from $^{56}$Ni by one- and two $\alpha$-particles.
For convenience, a linear  in $A$ term has been added, in order to give the equal energies to
$^{52}$Fe and $^{60}$Zn \cite{Bohr2}.  As it is seen in Fig. 1 the spectrum of the relative ground state energies
of the $\alpha$-particle type nuclei around $^{56}$Ni is close to the equidistant one. For this reason we approximate an $a$-dependence
of the collective potential by the harmonic oscillator. Thus, the collective potential used below is presented
by the following expression
\begin{eqnarray}
\label{eq8}
V(\Delta,a)=\frac{1}{2}C_{\Delta}\left(\frac{\Delta^4_0}{\Delta^2}+\Delta^2\right)+\frac{1}{2}C_{\alpha}a^2.
\end{eqnarray}
\begin{figure}[tbh]
\centerline{\includegraphics[width=7.3cm]{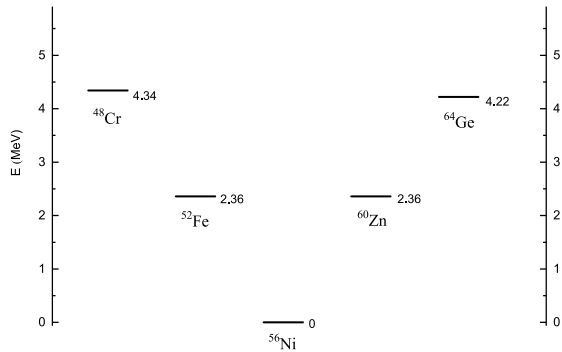}}
\caption{Experimental  relative energies of the ground states
of even-even nuclei that differ from $^{56}$Ni by one- and two $\alpha$-particles.
Energies are determined according to (\ref{eq6}). Linear in $(A-56)$ term is added
so as to give equal energies to $^{52}$Fe and $^{60}$Zn.
}
\label{Fig:1}
\end{figure}

\section{Results}

It is seen from (\ref{eq5}) that the pairing and $\alpha$-particle correlation modes are coupled
only through the pairing rotational term which can be presented for our aims as
\begin{eqnarray}
\label{eq9}
\frac{N^2}{B_{\Delta}\Delta^2+4B_{\alpha}a^2}=\frac{N^2}{4B_{\alpha}a^2}-\frac{N^2 B_{\Delta}\Delta^2}{4B_{\alpha}a^2(B_{\Delta}\Delta^2+4B_{\alpha}a^2)}.
\end{eqnarray}
Due to proportionality to $\Delta^2$ the second term in (\ref{eq9}) can be considered as a correction to Davidson
potential. As will be shown below this correction is small and can be neglected. However, we will retain this term by replacing
$\Delta^2$ and $a^2$ in the denominator of the second term on the right side of (\ref{eq9}) with their mean values $\langle\Delta^2\rangle$ and $\langle a^2\rangle$ in the
lowest states with given $N$ and $T$. As the result $\Delta$ and $a$ will be separated in the Hamiltonian. Thus, we obtain
\begin{eqnarray}
\label{eq10}
{\hat H}=\frac{T(T+1)}{2B_{\Delta}\Delta^2}+\frac{N^2}{2B_{\Delta}\Delta^2}
-\frac{1}{2B_{\Delta}\Delta^5}\frac{\partial}{\partial\Delta}\Delta^5\frac{\partial}{\partial\Delta}\nonumber\\
+\frac{1}{2}C_{\Delta}\frac{\Delta^4_0}{\Delta^2} +\frac{1}{2}C_{\Delta}\left(1-\frac{N^2B_{\Delta}}{4C_{\Delta}B_{\alpha}\langle a^2\rangle(B_{\Delta}\langle\Delta^2\rangle+4B_{\alpha}\langle a^2\rangle)}\right)\Delta^2\nonumber\\
-\frac{1}{2B_{\alpha}}\frac{\partial^2}{\partial a^2}+\frac{1}{2}C_{\alpha}a^2.
\end{eqnarray}
The eigenfunctions of (\ref{eq10}) are:
\begin{eqnarray}
\label{eq11}
&&\Psi_{NTT_z}(\Delta, a,\phi,{\vec \psi})\sim \nonumber\\
&&\sqrt{\frac{2T+1}{8\pi^2}}D^T_{T_z 0}({\vec \psi})
\exp(\imath N\phi)\exp\left(-\frac{1}{2}\sqrt{\frac{{\tilde C}_{\Delta}}{C_{\Delta}}}\cdot\sqrt{B_{\Delta}C_{\Delta}}\Delta^2-\frac{1}{2}\sqrt{B_{\alpha}C_{\alpha}}a^2\right)\nonumber\\
&\times&\left((B_{\Delta}C_{\Delta})^{1/4}\Delta\right)^{\sqrt{T(T+1)+\rho^4_0+4}-2}\left(2\left(\frac{C_{\Delta}B_{\alpha}}{B_{\Delta}C_{\alpha}}\right)^{1/4}(B_{\alpha}C_{\alpha})^{1/4}a\right)^{\frac{1}{2}(1+\sqrt{1+N^2})}\nonumber\\
&\times& L_n^{(\sqrt{T(T+1)+\rho^4_0+4})}\left(\sqrt{\frac{{\tilde C}_{\Delta}}{C_{\Delta}}}\cdot\sqrt{B_{\Delta}C_{\Delta}}\Delta^2\right)\times L_k^{(\frac{1}{2}\sqrt{1+N^2})}\left(\sqrt{B_{\alpha}C_{\alpha}}a^2\right). 
\end{eqnarray}
Here $L_{n(k)}$ are Laguerre polinomials, where $n$ and $k$ are positive integer numbers.
For each set of $N$ and $T$ for the lowest states $n=k=0$. Also,  $\rho_0^4\equiv B_{\Delta}C_{\Delta}\Delta_0^4$ and
\begin{eqnarray}
\frac{{\tilde C}_{\Delta}}{C_{\Delta}}=1-\frac{N^2B_{\Delta}}{4C_{\Delta}B_{\alpha}\langle a^2\rangle(B_{\Delta}\langle\Delta^2\rangle+4B_{\alpha}\langle a^2\rangle)}\nonumber
\end{eqnarray}
Using the wave functions (\ref{eq11}) we can calculate  $\langle N,T|\Delta^2|N,T\rangle$ and
$\langle N,T|a^2|N,T\rangle$. The results are
\begin{eqnarray}
\label{eq12}
\langle N,T|\Delta^2|N,T\rangle =\frac{C_{\Delta}}{{\tilde C}_{\Delta}}\frac{\left(\sqrt{T(T+1)+\rho_0^4+4}\right)}{\sqrt{B_{\Delta}C_{\Delta}}},
\end{eqnarray}
\begin{eqnarray}
\label{eq13}
\langle N,T|a^2|N,T\rangle=\frac{\left(3+\sqrt{1+N^2}\right)}{2\sqrt{B_{\alpha}C_{\alpha}}}.
\end{eqnarray}
With the help of (\ref{eq12}) and (\ref{eq13}) we obtain the following relation for ${\tilde C}_{\Delta}$/$C_{\Delta}$:
\begin{eqnarray}
\label{eq14}
&\displaystyle{\frac{{\tilde C}_{\Delta}}{C_{\Delta}}}& = \\
1 &-&\frac{N^2}{4R^2(3+\sqrt{1+N^2})^2+2R(\frac{{\tilde C}_{\Delta}}{C_{\Delta}})^{-1/2}(3+\sqrt{1+N^2})\sqrt{T(T+1)+\rho_0^4+4}}, \nonumber
\end{eqnarray}
where $R=\sqrt{(C_{\Delta}B_{\alpha})/(B_{\Delta}C_{\alpha})}$. Using the values of the parameters fixed below we obtain that even for $N$=4 ${\tilde C}_{\Delta}/C_{\Delta}\ge$ 0.99.

For the lowest states with given $N$ and $T$ the energies calculated from the energy of the basic nucleus having $N$=0, $T$=0 are
given by the expression
\begin{eqnarray}
\label{eq15}
E(N,T)&=&\frac{1}{2}\sqrt{\frac{C_{\alpha}}{B_{\alpha}}}\left(\sqrt{1+N^2}-1\right)\nonumber\\
&+&\sqrt{\frac{C_{\Delta}}{B_{\Delta}}}\left(\sqrt{T(T+1)+\rho^4_0+4}-\sqrt{\rho^4_0+4}\right).
\end{eqnarray}
The wave functions and the energies $E(N,T)$ depend on two dimensional
$\sqrt{C_{\Delta}/B_{\Delta}}$ and $\sqrt{C_{\alpha}/B_{\alpha}}$ and one nondimensional
$\rho_0$ parameters. As in our previous paper, we will compare with the experimental data
the ratios of the energies determined by (\ref{eq15}). To reduce the number of
parameters for calculating these quantities we fix the relative energy of the nucleus with
$N$=1 and $T$=1, whose experimental value is 4.5 MeV. It gives the relation between
$\sqrt{C_{\Delta}/B_{\Delta}}$, $\sqrt{C_{\alpha}/B_{\alpha}}$ and $\rho_0$
\begin{eqnarray}
\label{eq16}
\sqrt{\frac{C_{\Delta}}{B_{\Delta}}}\left(\sqrt{\rho^4_0+6}-\sqrt{\rho^4_0+4}\right)+\frac{1}{2}\sqrt{\frac{C_{\alpha}}{B_{\alpha}}}\left(\sqrt{2}-1\right)=4.5 \, {\rm MeV}.
\end{eqnarray}
The value of $\sqrt{C_{\alpha}/B_{\alpha}}$ is fixed as follows. Consider $0^+$ excited states
of nuclei around $^{56}$Ni having 4p-4h structure which can be considered as a quartet type excitations.
There are known $0^+$ state in $^{40}$Ca with the excitation energy $E^*$=3.35 MeV, $0^+$ state in $^{48}$Ca with  $E^*$=4.28 MeV,
$0^+$ state in $^{56}$Ni with  $E^*$=3.96 MeV,
$0^+$ state in $^{58}$Ni with  $E^*$=2.94 MeV, and $0^+$ state in $^{56}$Fe with  $E^*$=2.56 MeV.
The excitation energies in these nuclei vary from 2.5 MeV to 4.0 MeV with the average
value 3.4 MeV. We take as an approximate estimation of this energy the value of
$\sqrt{C_{\alpha}/B_{\alpha}}$  which is an energy of the $\alpha$-particle
type oscillations in harmonic approximation. This makes it possible to establish the relation
between $\sqrt{\frac{C_{\Delta}}{B_{\Delta}}}$ and $\rho_0$. We used in calculations two values of
$\sqrt{C_{\alpha}/B_{\alpha}}$, namely, 3 MeV and 4 MeV. The results of calculations with these two
values did not differ significantly, but nevertheless better agreement with the
experimental data was obtained with $\sqrt{C_{\alpha}/B_{\alpha}}$=3 MeV and
$\rho_0^4$=6. The results of calculations with these values of the parameters
and their comparison with the experimental data are presented in Fig. 2, where the energies
of the states with different $N$ and $T$ are counted from the energy of the state with $N=0, T=0$
and are given in units of $\left(E(N=1,T=1)-E(N=0,T=0)\right)$. As it is seen from Fig. 2
the results of the present calculations of the energies of the $T$=0 states are in a quite
satisfactory agreement with the experimental data. Comparison with the results obtained
in \cite{Jolos2} demonstrates how important is the inclusion of the $\alpha$-particle type correlations
in description of the $T$=0 states. The agreement with the experimental energies of the states with $T\ne 0$
is also satisfactory. This improvement of the results of the present calculations
in comparison with the previous ones \cite{Jolos2} for the states with different $T$
is explained by inclusion of  the $\alpha$-particle type correlations into consideration.
Due to this fact it becomes possible to fit the value of $\rho_0$ so as to get
a better description of the energies of the states with $T\ne 0$ without worsening description
of the energies of the states with $T$=0.
\begin{figure}[tbh]
\centerline{\includegraphics[width=7.3cm]{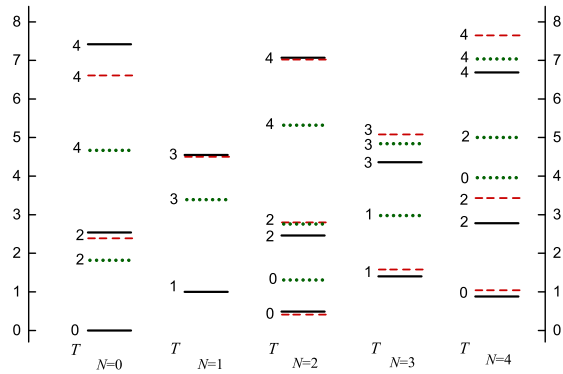}}
\caption{Experimental and calculated relative energies of the ground states
of even-even nuclei with different $N$ and $T$.
Thick solid line
(black) - experimental data, dashed line (red) - the results of calculations
with $\rho^4_0$=6 and $\sqrt{C_{\alpha}/B_{\alpha}}$= 3 MeV, dotted line (green) - the results
of calculations without inclusion of the $\alpha$-particle type correlations, taken from \cite{Jolos2}.
Energies are given in units of $\left(E(N=1,N=1)-E(N=0,T=0)\right)$.}
\label{Fig:2}
\end{figure}

\section{Two-nucleon and $\alpha$-particle transfer reactions}

The wave functions of the lowest states with given $N$ and $T$ are
\begin{eqnarray}
\label{eq17}
\Psi_{NTT_z}(\Delta, a,\phi,{\vec \psi})\sim\sqrt{\frac{2T+1}{8\pi^2}}D^T_{T_z 0}({\vec \psi})
\exp(\imath N\phi)\nonumber\\
\times\exp\left(-\frac{1}{2}\sqrt{B_{\Delta}C_{\Delta}}\Delta^2-\frac{1}{2}\sqrt{B_{\alpha}C_{\alpha}}a^2\right)\nonumber\\
\times\left((B_{\Delta}C_{\Delta})^{1/4}\Delta\right)^{\sqrt{T(T+1)+\rho^4_0+4}-2}\left((B_{\alpha}C_{\alpha})^{1/4}a\right)^{\frac{1}{2}(1+\sqrt{1+N^2})}
\end{eqnarray}
The operator corresponding to the two-nucleon transfer \cite{NPA183} is
\begin{eqnarray}
\label{eq18}
P_{\mu}^{(\pm 1)}=\pm\imath\exp(\pm\imath\phi)\Delta D^1_{\mu 0}
\end{eqnarray}
The operator associated with the $\alpha$-particle transfer is given by \cite{Dussel1}
\begin{eqnarray}
\label{eq18a}
S^{(\pm)}=\exp(\pm 2\imath\phi)a,
\end{eqnarray}
where it is assumed that $\alpha$-particle transfer is described by the $\alpha$-particle mode only.
Using (\ref{eq17}) and (\ref{eq18}) we obtain the following result
for the two-nucleon transfer amplitude
\begin{eqnarray}
\label{eq19}
\langle N+1,T'T_z'|P_{\mu}^{(+1)}|NTT_z\rangle=(B_{\Delta}C_{\delta})^{-1/4}C^{T'T_z'}_{TT_z  1\mu}
C^{T'0}_{T0 10}\nonumber\\
\times\frac{\Gamma\left(1+\frac{1}{2}\sqrt{T'(T'+1)+\rho_0^4+4}+\frac{1}{2}\sqrt{T(T+1)+\rho_0^4+4}\right)}
{\sqrt{\Gamma\left(\frac{1}{2}+\sqrt{T'(T'+1)+\rho_0^4+4}\right)\Gamma\left(\frac{1}{2}+\sqrt{T(T+1)+\rho_0^4+4}\right)}}\nonumber\\
\times\frac{\Gamma\left(\frac{3}{2}+\frac{1}{4}\sqrt{1+N^2}+\frac{1}{4}\sqrt{1+(N+1)^2}\right)}
{\sqrt{\Gamma\left(\frac{3}{2}+\frac{1}{2}\sqrt{1+N^2}\right)\Gamma\left(\frac{3}{2}+\frac{1}{2}\sqrt{1+(N+1)^2}\right)}}
\end{eqnarray}

It is seen from (\ref{eq19}) that the amplitude for the $\Delta T=0$ transition
is equal to zero if excitations in $\theta$-mode are excluded from consideration.
In this approximation the reaction $\left(^3He,p\right)$
to a state of an odd-odd nucleus with the same $T$ is forbidden if even-even
nucleus is a target \cite{NPA183}.

For the $\alpha$-particle transfer amplitude we obtain
\begin{eqnarray}
\label{eq20}
\langle N+2,T T_z|S^{(+)}_{\alpha}|NTT_z\rangle=\nonumber\\
\left(B_{\alpha}C_{\alpha}\right)^{-1/4}
\frac{\Gamma\left(2+\frac{1}{4}\sqrt{1+N^2}+\frac{1}{4}\sqrt{1+(N+2)^2}\right)}
{\sqrt{\Gamma\left(\frac{3}{2}+\frac{1}{2}\sqrt{1+N^2}\right)\Gamma\left(\frac{3}{2}+\frac{1}{2}\sqrt{1+(N+2)^2}\right)}}
\end{eqnarray}
This amplitude does not depend on isospin.

\section{Conclusion}

In the present paper we have constructed the collective Hamiltonian which includes not only the isovector pairing mode, but also the $\alpha$-particle degree of freedom describing the $\alpha$-particle
type correlations. The Hamiltonian is applied to  description of the relative energies
of the ground states of even-even nuclei around $^{56}$Ni. The results obtained demonstrate a quite satisfactory agreement with the experimental energies especially for the states
with $T$=0, i.e. for nuclei which are systems of some numbers of $\alpha$-particles. The agreement with the experimental energies of the states with $T\ne 0$ is also satisfactory.
This improvement of the results of calculations in comparison with the previous ones \cite{Jolos2}
is explained by inclusion in the Hamiltonian of the dynamical variable describing $\alpha$-particle type correlations. Because of this it becomes possible to include into consideration a difference
in description of the moments of inertia for isospin and gauage rotations.

\section*{Acknowledgements}

The authors acknowledge the partial support from
the Heisenberg-Landau Program. RVJ acknowledge support by the Ministry
of Education and Science (Russia) under Grant No. 075-10-2020-117.
EAK acknowledge support by Russian Foundation for Basic Research
under Grant No. 20-02-00176.

\appendix
\section{Appendices}

The classical expression for the kinetic energy looks as
\begin{eqnarray}
\label{eqa1}
T=\frac{1}{2}B_{\Delta}\sum_{\mu}{\dot z}^*_{1\mu}{\dot z}_{1\mu}+\frac{1}{2}B_{\partial}{\dot \alpha}^*{\dot \alpha},
\end{eqnarray}
where
\begin{eqnarray}
\label{eqa2}
z_{1\mu}=\exp(\imath \phi)\sum_{\nu}D^1_{\mu \nu}\Delta_{\nu},
\end{eqnarray}
\begin{eqnarray}
\label{eqa3}
\Delta_0=\Delta\cos\theta,\quad \Delta_{\pm 1}=\frac{1}{\sqrt{2}}\Delta\sin\theta,\nonumber\\
\alpha=\exp(2\imath\phi)a,
\end{eqnarray}
\begin{eqnarray}
\label{eqa4}
{\dot z}_{1\mu}=\exp(\imath \phi)\sum_{\nu}{\dot D}^1_{\mu \nu}\Delta_{\nu}+\imath{\dot \phi}\exp(\imath \phi)\sum_{\nu}D^1_{\mu \nu}\Delta_{\nu}+\exp(\imath \phi)\sum_{\nu}D^1_{\mu \nu}{\dot \Delta}_{\nu},
\end{eqnarray}
\begin{eqnarray}
\label{eqa5}
{\dot \alpha}=2\imath{\dot \phi}\exp(2\imath\phi)a + \exp(2\imath\phi){\dot a}.
\end{eqnarray}
Using the expression for ${\dot D}^1_{\mu \nu}$ \cite{Greiner} we obtain for (\ref{eqa4})
\begin{eqnarray}
\label{eqa6}
{\dot z}_{1\mu}=-\imath\exp(\imath \phi)\sum_{j=1,2,3}\sum_{\mu'} D^1_{\mu \mu'}\langle 1\mu'|{\hat L}_j|1\nu\rangle \omega_j\Delta_{\nu}\nonumber\\
+\imath{\dot \phi}\exp(\imath \phi)\sum_{\nu}D^1_{\mu \nu}\Delta_{\nu}+\exp(\imath \phi)\sum_{\nu}D^1_{\mu \nu}{\dot \Delta}_{\nu}.
\end{eqnarray}
In (\ref{eqa6}) $\omega_j$ is the $j$th component of the angular velocity in isospace and
$\langle 1\mu'|{\hat L}_j|1\nu\rangle$ is the matrix element in the three-dimensional
representation of the $j$th component of the isospace angular momentum.

Substituting (\ref{eqa5}) and  (\ref{eqa6}) into (\ref{eqa1}) we obtain the following
expression for $T$:
\begin{eqnarray}
\label{eqa7}
T=T_{rot}+T'+T_{\alpha},
\end{eqnarray}
where
\begin{eqnarray}
\label{eqa8}
T_{rot}=\frac{1}{2}\sum_{k,j=1,2,3}\Im_{kj}\omega_k\omega_j,
\end{eqnarray}
\begin{eqnarray}
\label{eqa9}
\Im_{kj}=B_{\Delta}\sum_{\mu,\nu,\nu'}\langle 1\mu|{\hat L}_k|1\nu\rangle^*\langle 1\mu|{\hat L}_j|1\nu'\rangle\Delta_{\nu}\Delta_{\nu'},
\end{eqnarray}
\begin{eqnarray}
\label{eqa10}
T'=-\frac{1}{2}B_{\Delta}\sum_j\sum_{\mu,\nu,\nu'}D^1_{\mu\mu'}\langle 1\mu|{\hat L}_j|1\nu\rangle\omega_j\Delta_{\nu}\nonumber\\
\times\left({\dot \phi}\sum_{\nu'}D^1_{\mu\nu'}\Delta_{\nu'}-\imath\sum_{\nu'}D^1_{\mu\nu'}{\dot \Delta}_{\nu'}\right)\nonumber\\
\times D^1_{\mu\mu'}\langle 1\mu'|{\hat L}_j|1\nu'\rangle\omega_j\Delta_{\nu'}\nonumber\\
+\frac{1}{2}B_{\Delta}\sum_{\mu}\left({\dot \phi}\sum_{\nu}D^{1}_{\mu\nu}*\Delta_{\nu}+\imath\sum_{\nu}D^{1}_{\mu\nu}*{\dot \Delta}_{\nu}\right)\nonumber\\
\times\left({\dot \phi}\sum_{\nu}D^{1}_{\mu\nu}*\Delta_{\nu}+\imath\sum_{\nu}D^{1}_{\mu\nu}*{\dot \Delta}_{\nu}\right),
\end{eqnarray}
\begin{eqnarray}
\label{eqa11}
T_{\alpha}=\frac{1}{2}B_{\alpha}\left({\dot \phi}^2a^2+{\dot a}^2\right).
\end{eqnarray}
Using the standard expressions for the matrix elements of the isospin momentum operators and (\ref{eqa2}) we obtain 
\begin{eqnarray}
\label{eqa12}
\Im_{kj}=\delta_{kj}\Im_j,\nonumber\\
\Im_j=B_{\Delta}\sum_{\nu\nu'}\langle 1\nu |{\hat L}^2_j |1\nu'\rangle\Delta_{\nu}\Delta_{\nu'},
\end{eqnarray}
and finally
\begin{eqnarray}
\label{eqa13}
\Im_1=B_{\Delta}\Delta^2,\nonumber\\
\Im_2=B_{\Delta}\Delta^2\cos^2\theta,\nonumber\\
\Im_1=B_{\Delta}\Delta^2\sin^2\theta.
\end{eqnarray}
Thus,
\begin{eqnarray}
\label{eqa14}
T_{rot}=\frac{1}{2}B_{\Delta}\Delta^2\omega_1^2+\frac{1}{2}B_{\Delta}\Delta^2\cos^2\theta\omega_2^2+\frac{1}{2}B_{\Delta}\Delta^2\sin^2\theta\omega_3^2.
\end{eqnarray}
In a similar way we obtain for $T'$ the following expression
\begin{eqnarray}
\label{eqa15}
T'=-\sqrt{2}B_{\Delta}{\dot \phi}\omega_1\Delta^2\sin2\theta +\frac{1}{2}B_{\Delta}\Delta^2{\dot \phi}^2+\frac{1}{2}B_{\Delta}\left({\dot \Delta}^2+\Delta^2{\dot \theta}^2\right)
\end{eqnarray}
For $T_{\alpha}$ we get
\begin{eqnarray}
\label{eqa16}
T_{\alpha}=\frac{1}{2}B_{\alpha}\left(4{\dot \phi}^2a^2+{\dot a}^2\right).
\end{eqnarray}
Thus, a classical expression for $T$ is
\begin{eqnarray}
\label{eqa17}
T=\frac{1}{2}B_{\Delta}\Delta^2\omega_1^2+\frac{1}{2}B_{\Delta}\Delta^2\cos^2\theta\omega_2^2+\frac{1}{2}B_{\Delta}\Delta^2\sin^2\theta\omega_3^2\nonumber\\
-\frac{1}{2}\sqrt{2}B_{\Delta}\Delta^2\sin2\theta\left(\omega_1\omega_1+\omega_1\omega_1\right)\nonumber\\
+\frac{1}{2}\left(B_{\Delta}\Delta^2+4B_{\alpha}a^2\right){\dot \phi}^2\nonumber\\
+\frac{1}{2}B_{\Delta}\left({\dot \Delta}^2+\Delta^2{\dot \theta}^2\right)+\frac{1}{2}B_{\alpha}{\dot a}^2.
\end{eqnarray}
We can present (\ref{eqa17}) as follows
\begin{eqnarray}
\label{eqa18}
T=\frac{1}{2}\sum_{j,k}g_{jk}(\xi){\dot \xi}_j{\dot \xi}_k,
\end{eqnarray}
where $\xi_1$=$\beta_1$,  $\xi_2$=$\phi$, $\xi_3$=$\beta_2$, $\xi_4$=$\beta_3$, $\xi_5$=$\Delta$, $\xi_6$=$\theta$, and $\xi_7$=$a$.
Here $\beta_j$ is the angle characterizing rotation around $j$th axis in isospace.
In agreement with Pauli prescription the quantum expression for $T$ is \cite{Pauli,Greiner}
\begin{eqnarray}
\label{eqa19}
{\hat T}=-\frac{1}{2}\sum_{j,k}g^{-1/2}\frac{\partial}{\partial\xi_k}g^{1/2}\left(g^{-1}\right)_{kj}\frac{\partial}{\partial\xi_j},
\end{eqnarray}
where $g$ is determinant of the matrix $g_{jk}$, and $g^{-1}_{kj}$ is the inverse of the matrix
$g_{jk}$.

The elements of the matrix $g_{jk}$ are presented in (\ref{eqa17}). Its determinant $g$ is equal to
\begin{eqnarray}
\label{eqa20}
g=\frac{1}{16}B_{\Delta}^5\Delta^8\frac{\sin^24\theta}{\cos^22\theta}B_{\alpha}\left(B_{\Delta}\Delta^2\cos^22\theta+4B_{\alpha}a^2\right).
\end{eqnarray}
Further, using a relation between the components of the isospin momentum operators ${\hat T}_j$
and the derivatives over $\beta_j$:
\begin{eqnarray}
{\hat T}_j=-\imath\frac{\partial}{\partial\beta_j}\nonumber
\end{eqnarray}
we obtain the following expression for ${\hat T}$
\begin{eqnarray}
\label{eqa21}
{\hat T}=\frac{1}{2}\left(\frac{1}{\sqrt{g}}{\hat T}_x\sqrt{g}\frac{B_{\Delta}\Delta^2+4B_{\alpha}a^2}{\left(B_{\Delta}\Delta^2\cos^22\theta+4B_{\alpha}a^2\right)B_{\Delta}\Delta^2}{\hat T}_x\right.\nonumber\\
\left.+\frac{1}{\sqrt{g}}{\hat T}_y\sqrt{g}\frac{1}{B_{\Delta}\Delta^2\cos^22\theta}{\hat T}_y+\frac{1}{\sqrt{g}}{\hat T}_z\sqrt{g}\frac{1}{B_{\Delta}\Delta^2\sin^2\theta}{\hat T}_z\right.\nonumber\\
\left.+\frac{1}{\sqrt{g}}{\hat T}_x\sqrt{g}\frac{\sin2\theta}{B_{\Delta}\Delta^2\cos^22\theta+4B_{\alpha}a^2}{\hat N}\right.\nonumber\\
\left.+\frac{1}{\sqrt{g}}{\hat N}\sqrt{g}\frac{\sin2\theta}{B_{\Delta}\Delta^2\cos^22\theta+4B_{\alpha}a^2}{\hat T}_x\right.\nonumber\\
\left.+\frac{1}{\sqrt{g}}{\hat N}\sqrt{g}\frac{\sin2\theta}{B_{\Delta}\Delta^2\cos^22\theta+4B_{\alpha}a^2}{\hat N}\right.\nonumber\\
-\left.\frac{1}{\sqrt{g}}\frac{\partial}{\partial\Delta}\sqrt{g}\frac{1}{B_{\Delta}}\frac{\partial}{\partial\Delta}-\frac{1}{\sqrt{g}}\frac{\partial}{\partial\theta}\sqrt{g}\frac{1}{B_{\Delta}\Delta^2}\frac{\partial}{\partial\theta}\right.
\left.-\frac{1}{\sqrt{g}}\frac{\partial}{\partial a}\sqrt{g}\frac{1}{B_{\alpha}}\frac{\partial}{\partial a}\right).
\end{eqnarray}
Substituting the expression for $g$ from (\ref{eqa21}) and taking into account the assumption made above that $B_{\Delta}$ and $B_{\alpha}$ don't depend on dynamical variables, we obtain
\begin{eqnarray}
\label{eqa22}
{\hat T}&=&\frac{1}{2B_{\Delta}\Delta^2}\frac{B_{\Delta}\Delta^2+4B_{\alpha}a^2}{\left(B_{\Delta}\Delta^2\cos^22\theta+4B_{\alpha}a^2\right)}{\hat T}_x^2\nonumber\\
&+&\frac{1}{2B_{\Delta}\Delta^2\cos^2\theta}{\hat T}_y^2+\frac{1}{2B_{\Delta}\Delta^2\sin^2\theta}{\hat T}_z^2\nonumber\\
&+&\frac{\sin2\theta}{B_{\Delta}\Delta^2\cos^22\theta+4B_{\alpha}a^2}{\hat T}_x{\hat N}\nonumber\\
&+&\frac{1}{2}\frac{{\hat N}^2}{B_{\Delta}\Delta^2+4B_{\alpha}a^2}\nonumber\\
&-&\frac{1}{2\sqrt{B_{\Delta}}\Delta^4}\frac{1}{\sqrt{B_{\Delta}\Delta^2\cos^22\theta+4B_{\alpha}a^2}}\frac{\partial}{\partial\Delta}\frac{\Delta^4\sqrt{B_{\Delta}\Delta^2\cos^22\theta+4B_{\alpha}a^2}}{\sqrt{B_{\Delta}}}\frac{\partial}{\partial\Delta}\nonumber\\
&-&\frac{\cos2\theta}{2\sqrt{B_{\Delta}}\Delta\sin4\theta\sqrt{B_{\Delta}\Delta^2\cos^22\theta+4B_{\alpha}a^2}}\frac{\partial}{\partial\theta}\frac{\sin4\theta\sqrt{B_{\Delta}\Delta^2\cos^22\theta+4B_{\alpha}a^2}}{\sqrt{B_{\Delta}}\Delta\cos2\theta}\frac{\partial}{\partial\theta}\nonumber\\
&-&\frac{\sqrt{B_{\Delta}}}{2B_{\alpha}\sqrt{B_{\Delta}\Delta^2\cos^22\theta+4B_{\alpha}a^2}}\frac{\partial}{\partial a}\frac{\sqrt{B_{\Delta}\Delta^2\cos^22\theta+4B_{\alpha}a^2}}{\sqrt{B_{\Delta}}}\frac{\partial}{\partial a}.
\end{eqnarray}

\end{document}